\documentclass[%
  a4paper,
  amsfonts,amssymb,amsmath,
  reprint,
  superscriptaddress,        
  showkeys,nofootinbib,twoside
]{revtex4-2}
\usepackage[english]{babel}
\usepackage[utf8]{inputenc}
\usepackage{amsthm}
\usepackage{siunitx}
\usepackage[version=4]{mhchem}
\usepackage{comment}
\usepackage{graphicx}
\usepackage{subcaption}
\usepackage[colorinlistoftodos, color=green!40, prependcaption]{todonotes}
\usepackage{lineno}
\usepackage[pdftex, pdftitle={Article}, pdfauthor={Author}]{hyperref} 

\usepackage{ragged2e}
\begin{document}

\title{Single-Photon Nonlinearity from a Nanobeam with a Quantum Dot}

\author{Abhijit Biswas}
\email{abiswas3@umd.edu}
\affiliation{Institute for Research in Electronics and Applied Physics, Joint Quantum Institute, and Department of Electrical and Computer Engineering, University of Maryland, College Park, Maryland 20742, USA}

\author{Neelesh Kumar Vij}
\affiliation{Institute for Research in Electronics and Applied Physics, Joint Quantum Institute, and Department of Electrical and Computer Engineering, University of Maryland, College Park, Maryland 20742, USA}

\author{Allan S. Bracker}
\affiliation{Naval Research Laboratory, 4555 Overlook Avenue SW, Washington, D.C. 20375, USA}

\author{Margaret Stevens}
\affiliation{Naval Research Laboratory, 4555 Overlook Avenue SW, Washington, D.C. 20375, USA}

\author{Joel Q. Grim}
\affiliation{Naval Research Laboratory, 4555 Overlook Avenue SW, Washington, D.C. 20375, USA}

\author{Edo Waks}
\affiliation{Institute for Research in Electronics and Applied Physics, Joint Quantum Institute, and Department of Electrical and Computer Engineering, University of Maryland, College Park, Maryland 20742, USA}

\begin{abstract}

An efficient single-photon nonlinearity is a key resource for photonic quantum technologies. Single-photon nonlinearities have been demonstrated with quantum dots (QDs) across a range of cavity and waveguide geometries. However, none of these realizations simultaneously provide high-efficiency direct fiber coupling and compatibility with scalable on-chip integration. The nanobeam cavity addresses these limitations through its in-plane geometry, which enables highly efficient, direct fiber coupling alongside scalable homogeneous and heterogeneous on-chip integration. Here, we report a single-photon nonlinearity from a nanobeam cavity coupled to a single quantum dot. The nanobeam interfaces directly with a single-mode fiber with an efficiency of $60\%$. Driven by resonant picosecond pulses, the platform achieves a single-photon nonlinearity with a low threshold of $0.24$ incident photons per pulse. The system operates in the strong-coupling regime, exhibiting a coupling rate of $g/2\pi = 29.1\text{ GHz}$ and a cooperativity of $C = 3.98$. These results establish the nanobeam cavity-QD system as a compact, chip-integrable, and fiber-compatible building block for scalable photonic quantum technologies.

\end{abstract}

\maketitle


\section{\label{sec:level1}Introduction}

Single-photon nonlinearity enables strong interactions between individual photons. Any quantum photonic application that requires individual photons to interact relies on a device that can generate an efficient nonlinearity at the single-photon level. These include nonclassical light generation~\cite{Faraon2008, Reinhard2012}, quantum information processing~\cite{Ralph2015, Yang2022, Basani2025}, photon--photon logic gates~\cite{Reiserer2014}, efficient Bell-state analysis~\cite{Ralph2015}, and all-optical switching~\cite{Chang2014, Chang2007, Chen2013}. Researchers have demonstrated this nonlinearity across many physical systems, with canonical realizations using real atoms coupled to microresonators~\cite{Birnbaum2005, Kubanek2008, Dayan2008, Aoki2009, Tiecke2014, Volz2014, Shomroni2014, Rosenblum2016, Scheucher2016, Liu2023}, and optical fibers~\cite{Prasad2020}. Atom trapping, however, requires a complex apparatus that is impractical for scalable on-chip operation, as is the case for organic emitters~\cite{Wang2019}. This has driven the search for solid-state implementations. Circuit QED platforms deliver single-photon nonlinearities by exploiting the large anharmonicity of the superconducting artificial atoms~\cite{Lang2011, Hoi2012, Vaneph2018}. Similarly, quantum emitters in semiconductors such as quantum dots~\cite{Sun2020} and defect centers~\cite{Sipahigil2016, Bhaskar2017, Pasini2024} can mediate strong photon--photon interactions. In particular, semiconductor quantum dots (QDs) are exceptional candidates for realizing single-photon nonlinearity, owing to their large oscillator strength, fast emission, and scalable integration with solid-state platforms~\cite{Lodahl2015, Heindel2023}.

Single-photon nonlinearities have been realized using semiconductor quantum dots across a variety of photonic architectures. Specific implementations include microcavities~\cite{Loo2012, Snijders2016, DeSantis2017, Snijders2018, Najer2019, Antoniadis2022, Wu2023, Tomm2024, Wang2026, Srinivasan2007}, photonic-crystal nanocavities~\cite{Faraon2008b, Reinhard2012, Muller2015, Englund2007, Fushman2008, Englund2012}, and photonic-crystal waveguides~\cite{Bennett2016, Hallett2018, LeJeannic2021, LeJeannic2022}. Despite this progress, several persistent limitations remain: many implementations rely on cross-polarization post-selection to reject uncoupled laser light, while others suffer from low input and output coupling efficiencies due to out-of-plane coupling. A practical, broadly useful nonlinearity requires high coupling efficiency, operation without post-selection, high cooperativity, and seamless on-chip integration; some applications additionally demand direct coupling to an optical fiber~\cite{Uppu2021}. While recent low-reflectivity and near-1D-atom platforms circumvent polarization filtering, they fail to resolve the other constraints~\cite{DeSantis2017, Wang2026}. Photonic-crystal nanobeam cavities uniquely fulfill all of these demanding requirements. Nanobeam cavities combine an ultra-small mode volume with homogeneous and heterogeneous integration capability and direct, high-efficiency in-plane fiber coupling, making them highly successful as bright single-photon sources~\cite{Islam2024, Biswas2025, Aghaeimeibodi2018}. However, single-photon nonlinearity from a nanobeam coupled to a quantum dot has not yet been demonstrated.

In this work, we demonstrate single-photon nonlinearity from a nanobeam cavity coupled to a single quantum dot. We attain a single-photon nonlinearity with an ultra-low threshold of $0.24$ incident photons per pulse. Operating in the strong-coupling regime, our system achieves a cooperativity of $3.98$, and the nanobeam couples directly to an optical fiber with $60\%$ efficiency. These results establish this nanobeam cavity-QD system as a scalable, high-efficiency nonlinearity platform for integrated photonic quantum technologies.

\begin{figure*}[htbp]
    \centering
    \includegraphics[width=0.9\textwidth]{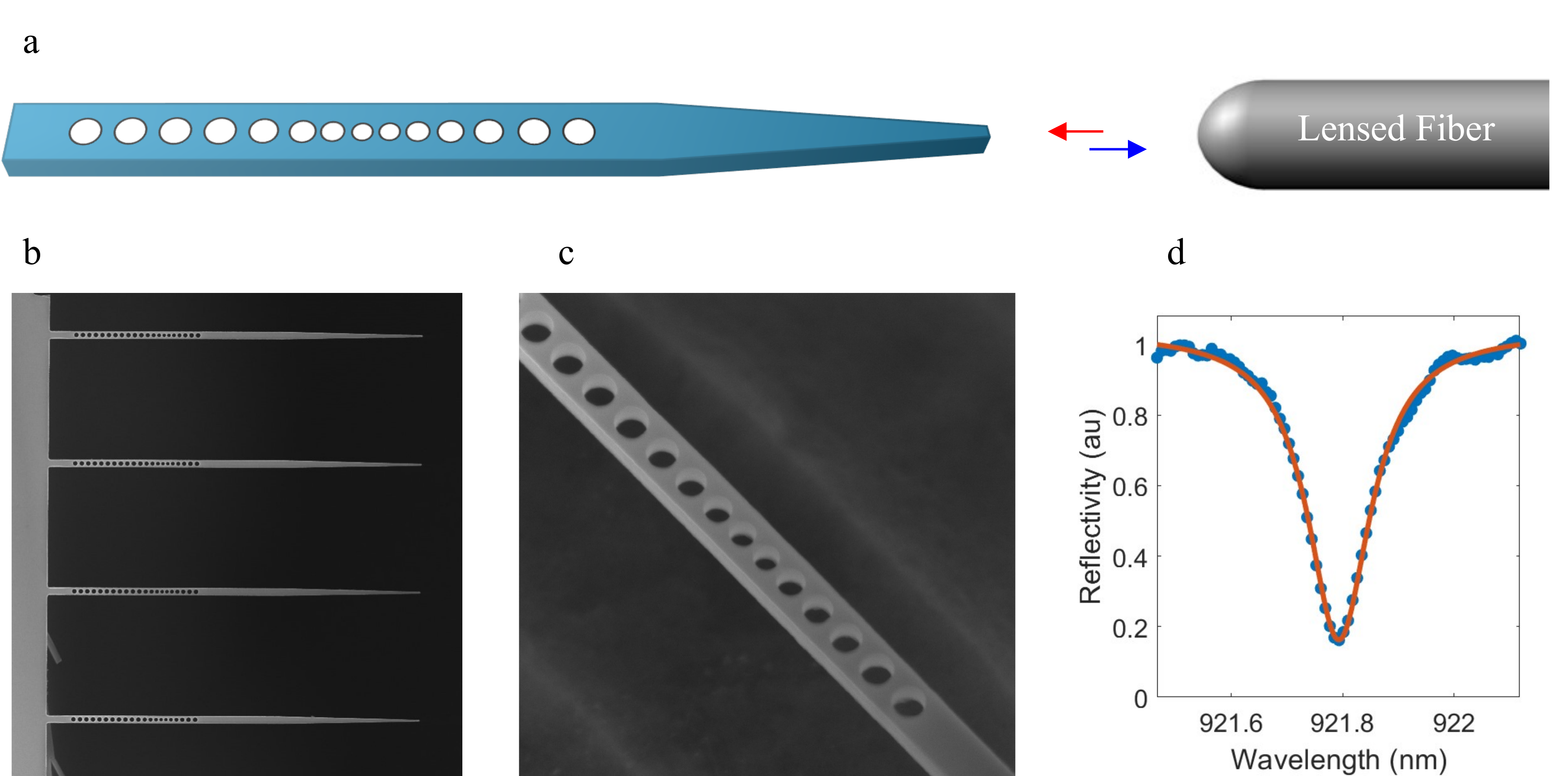}
    \caption{
        \textbf{Design and characterization of the nanobeam cavity.} 
        (a) Schematic of a nanobeam cavity with a tapered waveguide edge.
        (b) SEM image of suspended nanobeams from the top.
        (c) SEM image showing the cavity region.
        (d) Cavity reflectivity spectrum for cavity-mode characterization.
    }
    \label{fig:1}
\end{figure*}

\section{Design, Fabrication, and Characterization}

To achieve high-efficiency coupling to a fiber, we employ a tapered nanobeam~\cite{Deotare2009} structure, as shown in Fig.~\ref{fig:1}(a). This tapered edge minimizes spatial mode mismatch between the nanobeam emitted light and the lensed fiber mode, enabling direct, high-efficiency coupling. The photonic crystal cavity is formed by periodic air holes on a 1D air-clad nanobeam. Linearly tapering both the hole radii and lattice periodicity across a few central holes defines the cavity region. The cavity features an asymmetric design with fewer holes on the fiber-coupled side and a larger array on the opposite side, channeling cavity emission preferentially toward the nanobeam's tapered end.

Using three-dimensional finite-difference time-domain (FDTD) simulations, we optimized the nanobeam cavity parameters to achieve a high quality factor and an ultra-small mode volume at the quantum dot emission wavelength. We found that linearly tapering four central holes on each side of the cavity region yields optimal field confinement. For a symmetric structure with 15 mirror holes on both sides, simulations yield a quality factor of ${\sim}2\times 10^6$ and a mode volume of $0.1\times(\lambda/n)^3$, where $\lambda$ is the quantum dot emission wavelength. We systematically swept the hole count on the fiber-coupled side and found that six holes provide optimal optical access while preserving a sufficiently high quality factor. Additional details regarding the optimized parameters are provided in Supplementary Information.

\begin{figure*}[htbp]
    \centering
    \includegraphics[width=0.9\textwidth]{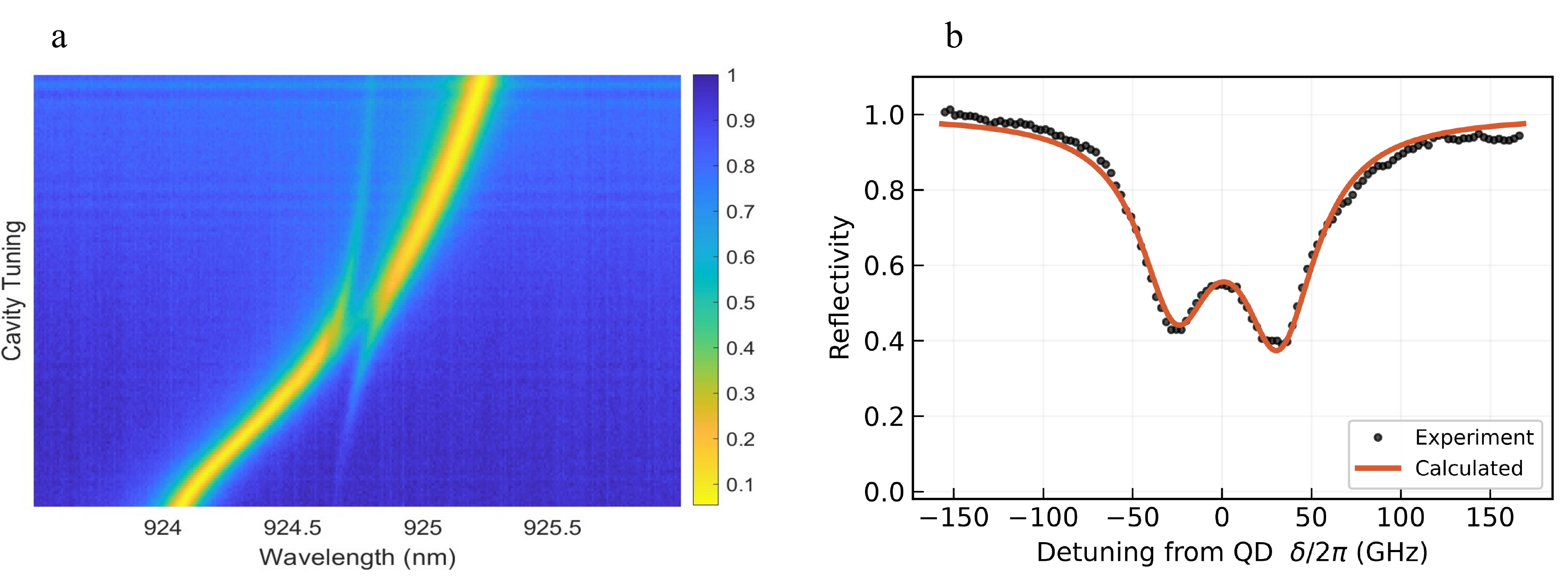}
    \caption{
        \textbf{Strong coupling between the cavity and the quantum dot.} 
        (a) Normalized cavity reflectivity under LED excitation while the cavity is tuned across the quantum dot emission wavelength.
        (b) Normalized cavity reflectivity with the quantum dot on resonance.
    }
    \label{fig:2}
\end{figure*}

The nanobeam devices were fabricated from a $190\text{ nm}$ thick GaAs membrane containing a single layer of self-assembled InAs quantum dots. We fabricated the nanobeam devices using standard e-beam lithography and reactive ion etching. Figures~\ref{fig:1}(b) and \ref{fig:1}(c) show SEM images of the completed devices. Additional fabrication procedures and wafer specifications are detailed in the Supplementary Information.

To enable direct optical access, we suspended the nanobeams along the chip edge via precision cleaving. Precision cleaving fractures the substrate such that the nanobeams remain attached on one side. This edge-suspension geometry facilitates rapid, stable, and highly efficient coupling to a lensed single-mode fiber. This approach is particularly advantageous for preserving electrical connectivity to the quantum dot through the doped layers in gated structures.

We performed the measurements using a fiber probe station at $5\text{ K}$ placed inside a closed-cycle cryostat. For resonant measurements, we excite the nanobeam through a lensed fiber. Above-band excitation is performed through an objective lens from the out-of-plane direction. We collect the nanobeam emission through the lensed fiber and route it through a 90:10 fiber beamsplitter for analysis with a spectrometer or photon detectors.

For initial cavity characterization, we reflected continuous-wave broadband light from a superluminescent light-emitting diode (SLED) off the nanobeam. Figure~\ref{fig:1}(d) shows the reflected intensity as a function of wavelength near the cavity resonance. On resonance, the cavity scatters light out of the nanobeam plane, producing a dip in the reflected signal. The blue dots represent reflected intensity, and the solid red line is the Lorentzian fit. From the fit, we extracted a cavity decay rate $\kappa_1/2\pi = 48.43\text{ GHz}$ and a quality factor $Q = 6700$. On resonance, the reflectivity dip almost completely vanishes, indicating we are near the critical-coupling regime, where the out-of-plane scattering rate equals the rate at which light couples to the nanobeam's backward-propagating mode.

To measure the coupling efficiency, we performed a reflectivity measurement using a narrowband continuous-wave laser. We excited the nanobeam outside the cavity dip and measured the input and reflected power using a fiber coupler. From the measured powers, and assuming symmetric coupling between the nanobeam and lensed fiber, we calculated a fiber-to-nanobeam coupling efficiency ($\eta_{\text{coupling}}$) of $60\%$. Efficiency can be further improved by optimizing the taper design and increasing the lens's numerical aperture.

\section{Result}

\begin{figure*}[htbp]
    \centering
    \includegraphics[width=0.9\textwidth]{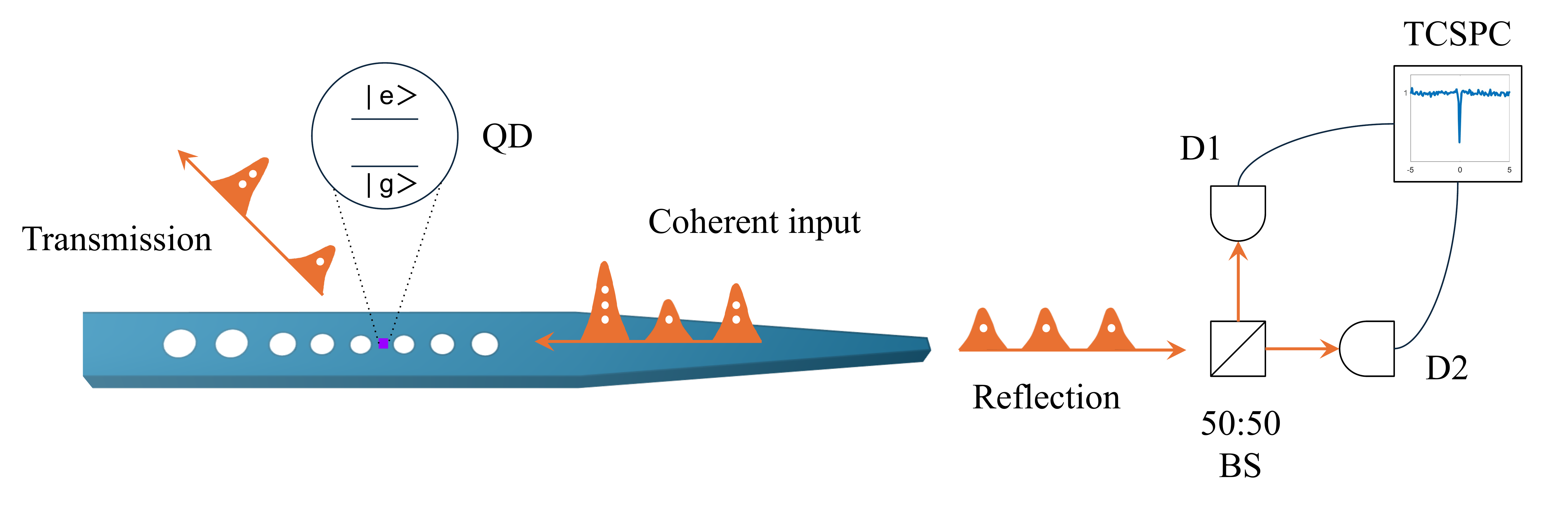}
    \caption{
        \textbf{Resonant reflection from the cavity-quantum dot system.} 
        At low photon number, the single-photon components of the incident light are preferentially reflected, while multi-photon components are transmitted.
    }
    \label{fig:3}
\end{figure*}

To identify a quantum dot coupled to the cavity, we tuned the cavity resonance while monitoring the reflectivity spectra under broadband SLED excitation. Figure~\ref{fig:2}(a) shows the reflectivity spectrum as a function of cavity tuning. We employed a gas-tuning method in which $\text{N}_2$ gas condenses locally on the nanobeam, causing a continuous red shift of the cavity resonance. When the cavity resonance is tuned across a coupled dot emission wavelength, a clear anticrossing is observed. This indicates a strong coupling between the cavity mode and the quantum dot. This approach lets us control the cavity--dot detuning and investigate coupling behavior over a range of detunings.

A quantum dot in a semiconductor often exhibits blinking due to charge fluctuations in the surrounding environment. To stabilize the emission for this identification measurement, we applied a weak above-band laser together with the SLED, which reduces charge fluctuations via optical gating and yields a more stable quantum-dot~\cite{Nguyen2012, Liu2018, Luo2019}. We observed that the same stabilization can be achieved using a collimated white-light source passing through a $900\text{ nm}$ low-pass filter. Owing to experimental constraints, we performed the single-photon-nonlinearity measurements described below without the above-band laser.

To determine the strength of the light--matter interaction, we analyzed the cavity reflectivity spectrum near resonance with the quantum-dot emission. Figure~\ref{fig:2}(b) shows the reflected intensity as a function of detuning from QD emission. The reflectivity shows a doublet corresponding to two polariton modes of the Jaynes-Cummings model. The solid line is a theoretical fit derived from the input-output formalism and the Jaynes-Cummings model (see Supplementary Information)~\cite{Luo2019}. From the fit, we found a cavity--quantum-dot coupling rate $g/2\pi = 29.1\text{ GHz}$ and a cavity-QD detuning of $5.95\text{ GHz}$.

The total quantum dot decay rate $\gamma$ influences the reflectivity, but we could not measure it reliably from the measured spectrum. In the theoretical model, $\gamma$ strongly influences the reflectivity peak at the cavity center, but because of quantum-dot blinking, the measured spectrum is a time-averaged response of both the bare cavity and the coupled system. Thus, the central peak does not accurately represent the coupled system alone. We therefore determine $\gamma$, together with the blinking fraction, from the power-dependent measurements presented below, and defer the cooperativity calculation to that analysis.

Optical access to the cavity is governed by the in-plane cavity decay rate ($\kappa_\parallel/2\pi$) into the collected nanobeam waveguide mode. The ratio of the in-plane decay rate to the total decay rate ($\kappa_\parallel/\kappa_1$) sets the cavity-nanobeam coupling condition. For critical coupling, $\kappa_\parallel/\kappa_1 = 0.5$. We determined this ratio from the minimum bare-cavity reflectivity using $R_{\min} = (1 - 2\kappa_\parallel/\kappa_1)^2$. With the cavity and quantum dot on resonance, we independently measured $\kappa_1$ and the ratio $\kappa_\parallel/\kappa_1$ using a high-power continuous-wave laser saturating the quantum dot. We found $\kappa_1/2\pi = 60.77\text{ GHz}$ and $\kappa_\parallel/\kappa_1$ to be either $0.35$ (undercoupled) or $0.65$ (overcoupled). We kept $\kappa_\parallel/\kappa_1$ as a fitting parameter in our theoretical model to unambiguously identify the actual coupling condition. The measured values satisfy $g > \kappa_1/4$, confirming that the system operates firmly within the strong-coupling regime.

To observe the single-photon nonlinearity, we excite the cavity with weak, resonant picosecond pulses. Figure~\ref{fig:3} illustrates the experiment. The excitation pulses follow a Poissonian photon-number distribution comprising predominantly single-, two-, and three-photon pulses. Upon entering the cavity, an initial photon is absorbed by the quantum dot, driving it to its excited state. Any subsequent photons arriving within the transition lifetime of the dot do not interact with it and instead transmit without interaction~\cite{Rosenblum2011}. They predominantly propagate into the out-of-plane modes. Once the quantum dot relaxes back to the ground state, it reflects a photon. The reflected signal is thus dominated by single-photon pulses, and the overall system can be viewed as a single-photon filter. The lensed fiber collects the reflected photons and routes them for correlation measurements.

To maximize photon--dot interaction, the photon pulse width should match the dot's transition lifetime. In this strongly coupled system, the dot lifetime is on the order of a few picoseconds ($1/(\kappa_1/2) \approx 6\text{ ps}$). A shorter pulse, however, carries a broader spectral bandwidth, placing a significant fraction of its energy outside the quantum-dot interaction bandwidth. We therefore chose an optimum $18\text{ ps}$ pulse as a compromise between temporal matching to the emitter lifetime and spectral matching to its bandwidth. We generated optical pulses by spectrally shaping the output of a mode-locked titanium--sapphire laser (pulse duration: $3\text{ ps}$, repetition rate: $f_{\text{rep}} = 76\text{ MHz}$).

To monitor the input power, we used a fiber coupler tap to continuously measure the power launched into the input optical fiber ($P_{\text{fiber}}$). The power coupled into the nanobeam waveguide is $P_{\text{beam}} = \eta_{\text{coupling}} P_{\text{fiber}}$, where $\eta_{\text{coupling}}$ is the fiber-to-waveguide coupling efficiency. We then calculated the average number of photons per pulse from the incident power using $\langle n \rangle = P/(f_{\text{rep}} \hbar \omega_{\text{laser}})$, where $P$ is the average power and $\hbar \omega_{\text{laser}}$ is the laser photon energy. To distinguish between incident and device-coupled photon fluxes, we define two characteristic photon numbers per pulse: the in-fiber photon number $\langle n \rangle_{\text{fiber}}$, calculated from $P_{\text{fiber}}$, and the device-coupled photon number arriving inside the nanobeam $\langle n \rangle_{\text{beam}}$, calculated from $P_{\text{beam}}$. All subsequent power-dependent results and photon-correlation measurements are referenced to these parameters.

\begin{figure*}[htbp]
    \centering
    \includegraphics[width=0.9\textwidth]{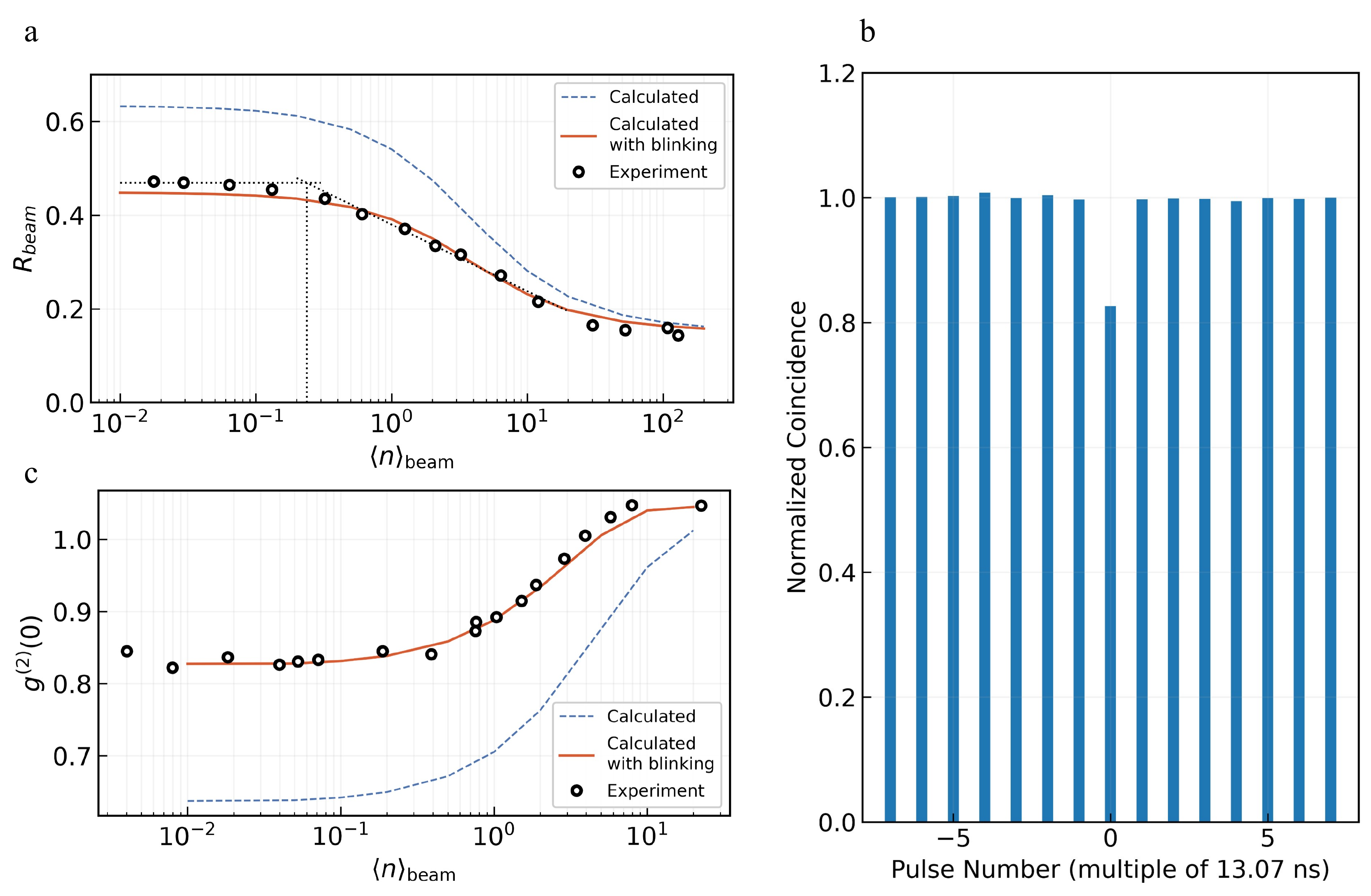}
    \caption{
        \textbf{Single-photon nonlinearity.} 
        (a) Reflectivity from the nanobeam as a function of $\langle n \rangle_{\text{beam}}$
        (b) Second-order autocorrelation measurement of the reflected signal for $\langle n \rangle_{\text{beam}}$ = 0.06.
        (c) Measured g2(0) as a function of incident photon number in the nanobeam, $\langle n \rangle_{\text{beam}}$.
    }
    \label{fig:4}
\end{figure*}

To determine the nanobeam's nonlinearity threshold, we measured its reflectivity across a range of input powers. We determined relative reflectivity by comparing the reflected signal at the photon detector with the input power. To reconstruct the absolute power-dependent response, we normalized the high-power experimental data to the bare-cavity reflectivity limit extracted from numerical simulations.

To model the power-dependent response, we solve the Jaynes--Cummings master equation for the resonantly driven quantum-dot--cavity system. We excite the dot with a pulse and propagate the emitted field to the detector via the input--output relation, retaining both the coherently reflected drive and the cavity-radiated field. We model quantum-dot blinking as a statistical mixture of the coupled and bare-cavity configurations. The simulation parameters---the cavity linewidth, vacuum Rabi splitting, detuning, and pulse duration---are fixed by reflectivity characterization and measurements. In contrast, we obtain the ratio $\kappa_\parallel/\kappa_1$, the total decay rate $\gamma$, and the blinking ratio $r$ by fitting the model to the power-dependent reflectivity and $g^{(2)}(0)$ measurements. The Supplementary Information provides additional details of the theoretical model.

Figure~\ref{fig:4}(a) shows the reflectivity as a function of the average photon number per pulse inside the nanobeam, $\langle n \rangle_{\text{beam}}$. The black circles are the experimental data, normalized to the bare-cavity reflectivity limit. The dashed curve shows the calculated reflectivity of the coupled quantum-dot--cavity system, and the solid curve shows the same calculation after accounting for quantum-dot blinking (with an extracted blinking ratio of $r = 0.62$). At very low excitation power, the reflectivity reaches approximately $47\%$, indicating strong interaction between the incident photons and the quantum dot. As the input power increases, the reflectivity decreases and eventually saturates at the bare-cavity value of around $15\%$, consistent with saturation of the quantum-dot transition and a reduced nonlinear response.

The nonlinearity threshold is defined as the intersection of the constant low-power plateau with a linear fit to the saturating roll-off (in $R$ versus $\log_{10} \langle n \rangle$ of Fig.~\ref{fig:4}(a)). We found the nanobeam's nonlinearity threshold to be $0.24$ photons per pulse. This value is of the same order as the state of the art on comparable solid-state platforms~\cite{DeSantis2017} and highlights the high efficiency of the quantum-dot--photon interaction in our system.

We measured the second-order correlation $g^{(2)}(\tau)$ of the reflected signal using a Hanbury Brown and Twiss (HBT) setup. We sent the reflected signal through a 50:50 beam splitter and detected it with two superconducting nanowire single-photon detectors (SNSPDs). The output was analyzed with a time-correlated single-photon counting (TCSPC) module. Figure~\ref{fig:4}(b) shows the measured normalized coincidence count as a function of delay at $\langle n \rangle_{\text{beam}} = 0.06$ photons per pulse. The reduced zero-delay peak $g^{(2)}(0) = 0.83$ confirms the presence of single-photon components.

The relatively high $g^{(2)}(0)$ can be attributed to a few factors. First, the cavity is not perfectly critically coupled, so some light is always reflected without any nonlinear interaction. Second, the laser's time--bandwidth product makes its temporally short pulses spectrally wide, so some pulse energy falls outside the quantum-dot bandwidth and is reflected without interacting with the dot~\cite{Sun2020, LeJeannic2022}. Because we performed these measurements without optical or electrical gating, the quantum dot was also subject to blinking due to an unstable charge environment. Stabilizing the charge environment---either with a weak above-band laser or, more robustly, by embedding the dot in a diode structure---would improve the performance.

We observe that the statistics of the scattered photons depend strongly on the incident photon flux. Figure~\ref{fig:4}(c) shows $g^{(2)}(0)$ as a function of the average photon number per pulse inside the nanobeam, $\langle n \rangle_{\text{beam}}$. The dashed curve corresponds to the model without blinking, and the solid curve represents the model incorporating quantum dot blinking. At low excitation level, the nonlinear interaction with the quantum dot strongly suppresses the multi-photon components of the reflected signal. The reflected light is dominated by single-photon pulses and $g^{(2)}(0)$ is reduced, indicating sub-Poissonian statistics. As $\langle n \rangle_{\text{beam}}$ increases, the quantum dot saturates, causing an increasing fraction of the pulse to reflect without nonlinear interaction. Consequently, the coherent component of the reflected field increases, driving the photon statistics toward a Poissonian distribution ($g^{(2)}(0) \to 1$). At the highest input photon numbers, $g^{(2)}(0)$ rises above unity, indicating weak super-Poissonian photon-bunching statistics, consistent with our Jaynes--Cummings model.

From the fit, we obtain a spontaneous decay rate $\gamma_{\text{sp}}/2\pi = 6\text{ GHz}$ and pure dephasing $\gamma^*/2\pi = 0.5\text{ GHz}$, giving a total decay rate $\gamma/2\pi = 3.5\text{ GHz}$ ($\gamma_{\text{sp}}/2 + \gamma^*$). The extracted spontaneous decay rate, $\gamma_{\text{sp}}/2\pi = 6\text{ GHz}$, is larger than values commonly reported for InAs quantum dots. We attribute this to the strong Purcell enhancement of the nanobeam cavity, consistent with the short lifetime ($\sim 6\text{ ps}$) estimated above. Comparably enhanced decay rates, well above the free-space value, have been reported for other strongly coupled quantum-dot--cavity systems, where the excess is attributed to Purcell acceleration and phonon-assisted emission into non-resonant modes~\cite{Loo2012, Ohta2011}. The low pure-dephasing rate $\gamma^*/2\pi = 0.5\text{ GHz}$ confirms that fast radiative decay, rather than decoherence, dominates this broadening.

With $\gamma$ determined, we calculate the cooperativity $C = g^2/\kappa_1\gamma = 3.98$. We also extract a cavity coupling ratio of $\kappa_\parallel/\kappa_1 = 0.6$, placing the nanobeam in the overcoupled regime, which agrees well within experimental uncertainty with our independently measured value of $0.65$.

Although the results in Figs.~\ref{fig:4}(a) and \ref{fig:4}(c) describe the standalone nanobeam device, the overall performance of the fiber-integrated system can be determined directly through an appropriate scaling. To transform the reflectivity response in Fig.~\ref{fig:4}(a) to the integrated-system frame, we scale the photon-number axis by $1/\eta_{\text{coupling}}$ and the reflectivity by $\eta_{\text{coupling}}^2$. In this integrated-system representation, we extract an effective fiber-to-fiber single-photon nonlinearity threshold of $0.40$ photons per pulse. For the second-order correlation in Fig.~\ref{fig:4}(c), we obtain the dependence on the in-fiber photon number $\langle n \rangle_{\text{fiber}}$ by rescaling the photon-number axis by $1/\eta_{\text{coupling}}$. As the reflected light couples back into the fiber with a fiber coupling efficiency of $\eta_{\text{coupling}}$, and linear loss does not alter the underlying photon statistics, the $g^{(2)}(0)$ values remain invariant to this output attenuation.

\section{Conclusion}
In summary, we demonstrated single-photon nonlinearity from nanobeam cavity coupled to a QD with a nonlinearity threshold as low as $0.24$ photons per pulse. The system operates in the strong coupling regime, with a coupling rate $g/2\pi = 29.1\text{ GHz}$ and a cooperativity $C = 3.98$, while achieving a direct fiber-coupling efficiency of $60\%$. The nanobeam geometry is largely material-agnostic, so it can equally host telecom-band quantum dots, whose performance and fabrication maturity are steadily improving~\cite{Vyvlecka2026, Albrechtsen2026, Kolatschek2021}, allowing chip- and fiber-compatible all-optical switches and photon--photon gates to be built directly at telecommunication wavelengths. Together, these capabilities establish the nanobeam cavity-QD paltform as a compact and scalable building block for photonic quantum technologies.

\section{Author Contribution}
A.B. designed and fabricated the device, built the optical measurement setups, performed the experiments, analyzed the data, carried out numerical modeling, and drafted the manuscript. N.K.V. assisted with numerical modeling. A.S.B., M.S., and J.Q.G. synthesized the quantum-dot-embedded wafers via molecular beam epitaxy. E.W. conceived the experiment, supervised the project, and revised the manuscript.

\bibliography{refs}

\end{document}


\maketitle
\vspace{-3.5em}

\section{Nanobeam cavity Design Parameters}
\label{sec:model}
Using FDTD simulations, we optimized the design for a high quality factor and ultra-small mode volume. The cavity resonance was targeted within the embedded InAs/GaAs quantum dot emission range, 910 $\sim$ 960 nm. The nanobeam has a width of 285 nm and a thickness of 190 nm. The period of the untapered mirror holes is 252 nm, and the radius is 83 nm. Near the cavity center, the period and radius are linearly tapered across 4 holes, gradually reaching 188 nm and 62 nm, respectively. The length of the adiabatic waveguide taper is 5.5 µm.

\section{Wafer and Fabrication Details}
\label{sec:model}
Self-assembled InAs quantum dots (QDs) were grown by molecular beam epitaxy (MBE) at the center of a GaAs membrane layer deposited above an AlGaAs sacrificial layer on a GaAs substrate.  The optimized fabrication process starts with spin-coating a GaAs chip with positive resist ZEP520A. We target a resist thickness of 500 nm. We patterned the nanobeam photonic cavity structures using an electron-beam lithography system. After following a cold-development method using o-xylene, the patterns were transferred to the GaAs sample using a chlorine-based ICP-RIE process. A wet etch using HF releases the nanobeams by removing the sacrificial layer. 

\section{Reflectivity Model}
\label{sec:model}
To determine the coupling rate between the cavity and QD, we fit the reflectivity spectrum to the following equation~\cite{Luo2019}
\begin{equation}
R = B + \frac{S}{\left| -i\Delta\omega + \frac{\kappa}{2} + \frac{g^2}{-i(\Delta\omega - \delta) + \gamma} \right|^2},
\label{eq:cavity_dot_R}
\end{equation}
where $g$ is the coupling rate between the cavity and the QD, $\gamma$ is the decay rate of the quantum dot, $\Delta\omega$ is the detuning from the cavity, and $\delta$ is the detuning between the quantum dot and the cavity. $B$ and $S$ represent fitting parameters denoting the baseline background signal and scaling factor, respectively.

\section{Theoretical Model}
\label{sec:model}

Throughout, we work in the frame rotating at the laser (equivalently QD)
frequency and adopt normalized units in which the total cavity energy-decay
rate is unity, $\kappa_1\equiv1$. All rates ($g_1$, $\gpar$, $\gdeph$,
$\kpar$), detunings ($\delta_{\mathrm{QD}}$, $\Delta$), times ($\sigma_t$,
$\Trep$), and the pulse bandwidth are expressed relative to $\kappa_1$; the
normalization is left implicit. We set $\hbar=1$.

\subsection{Jaynes--Cummings Hamiltonian with a pulsed drive}
The QD is modeled as a two-level emitter (lowering operator $\sm$, raising
operator $\sptt$, with $\sptt\sm=\ket{e}\!\bra{e}$ the excited-state
projector) coupled to a single cavity mode (annihilation operator $\aop$)
through the Jaynes--Cummings (JC) interaction~\cite{JaynesCummings1963,
ShoreKnight1993}. In the frame rotating at the drive frequency $\omega_L$ and
within the rotating-wave approximation, the Hamiltonian is
\begin{equation}
\hat{H}(t) = \hat{H}_0 + \hat{H}_{\mathrm{int}}
+ \hat{H}_{\mathrm{drive}}(t),
\label{eq:Htot}
\end{equation}
with 
\begin{align}
\hat{H}_0 &= \delta_{\mathrm{QD}}\,\sptt\sm + \Delta\,\ad\aop,
\label{eq:H0}\\
\hat{H}_{\mathrm{int}} &= g_1\!\left(\ad\sm + \aop\sptt\right),
\label{eq:Hint}\\
\hat{H}_{\mathrm{drive}}(t) &= \sqrt{\kpar}\eps(t)\!\left(\ad + \aop\right),
\label{eq:Hdrive}
\end{align}

\begin{equation}
\eps(t) = \sqrt{\ninbeam}~\!\left(\frac{4~\mathrm{ln}(2)}{\tau^2\pi}\right)^{1/4}~\exp\!\left[-2~\mathrm{ln}(2)\frac{(t-t_0)^2}{\tau^2}\right]
\label{eq:gauss}
\end{equation}

where $\delta_{\mathrm{QD}}$ and $\Delta$ are the QD-transition and cavity detunings
from the drive. The two detunings are retained as free parameters so that the treatment is general. In the present experiments the drive is
resonant with the QD, $\delta_{\mathrm{QD}}=0$ (laser at
$\lambda_{\mathrm{QD}}$), and the cavity carries the detuning
$\Delta=\omega_c-\omega_{\mathrm{QD}}$.
$g_1$ is the coherent coupling (half the vacuum-Rabi splitting),
$\kpar$ cavity coupling to the collection channel.
$\eps(t)$ is the real drive envelope and $\tau$ is the intensity temporal FWHM of the pulse. The pulse energy equals the mean input photon number inside the nanobeam device ($\int\eps(t)^2\,dt=\ninbeam$).
$\ninbeam$ is connected to the input photon number in fiber by the fiber-to-nanobeam coupling efficiency $\eta_\mathrm{coupling}$

\begin{equation}
\ninbeam = \eta_\mathrm{coupling}\nin
\label{eq:gauss}
\end{equation}

This coupling efficiency factor is not included in the drive Hamiltonian; instead, the input photon number was scaled and used in the drive term.

\subsection{Lindblad master equation}
Dissipation is treated with a Markovian master equation of Lindblad
form~\cite{Lindblad1976,GKS1976,BreuerPetruccione2002},
\begin{equation}
\dot{\hat\rho} = -i\,[\hat H(t),\hat\rho]
+ \mathcal{D}[\sqrt{\kappa_1}\,\aop]\hat\rho
+ \mathcal{D}[\sqrt{\gpar}\,\sm]\hat\rho
+ \mathcal{D}\!\left[\sqrt{\tfrac{\gdeph}{2}}\,\sz\right]\!\hat\rho,
\qquad
\mathcal{D}[\hat L]\hat\rho = \hat L\hat\rho\hat L^\dagger
- \tfrac12\{\hat L^\dagger\hat L,\hat\rho\},
\label{eq:me}
\end{equation}
with Liouvillian $\mathcal{L}$ defined by $\dot{\hat\rho}=\mathcal{L}\hat\rho$.
Three dissipative channels appear:
\begin{itemize}[leftmargin=1.6em,itemsep=0.15em]
\item total cavity photon loss $\sqrt{\kappa_1}\,\aop$;
\item QD \emph{population} (radiative) decay $\sqrt{\gpar}\,\sm$;
\item QD \emph{pure dephasing} $\sqrt{\gdeph/2}\,\sz$;
\end{itemize}

Each pulse begins in the product state
$\hat\rho_0=\ket{0}\!\bra{0}_{\mathrm{cav}}\otimes\ket{g}\!\bra{g}_{\mathrm{QD}}$
(empty cavity, QD in the ground state); all excitation is supplied by the
coherent drive $\hat H_{\mathrm{drive}}(t)$.

\subsection{Output field and detected flux}
\label{sec:output}
For a one-sided, coherently driven lossy cavity, input--output
theory~\cite{GardinerCollett1985,Collett1984} gives the detected output field
as a coherent (classical) reflected prefactor plus the radiated cavity field,
\begin{equation}
\aout(t) = \eps(t)\,\Id - i\sqrt{\kpar}\,\aop(t)
\label{eq:aout}
\end{equation}
where $\kpar\le\kappa_1$ is the rate at which intracavity light couples into
the collected mode, the remaining $\kappa_1-\kpar$ being lost through other channels.
The instantaneous output photon flux is
\begin{equation}
S_\mathrm{beam}(t) = \expval{\adout(t)\aout(t)}
\label{eq:S_of_t}
\end{equation}

The detected output photon flux is

\begin{equation}
S_\mathrm{fiber}(t) = \eta_\mathrm{coupling}S(t)
\label{eq:S_of_t}
\end{equation}

\subsection{Power-dependent reflectivity}
\label{sec:reflectivity}
The reflectivity is the ratio of output
to input pulse energy.
\begin{equation}
{\;
R_\mathrm{beam} = \frac{\Nbar_{\mathrm{out}}(\ninbeam)}{\ninbeam}
}
\label{eq:RvsP}
\end{equation}
where $\Nbar_{\mathrm{out}}=\int S_\mathrm{beam}(t)\,dt$ is the mean number of output photons.

\paragraph{Bare-cavity reference.} The saturated (high-power) response is the
bare-cavity reflectivity $\Rbare$, obtained from the identical calculation
with the QD decoupled, $g_1\to0$. Because the linear cavity is
power-independent, $\Rbare$ is evaluated once and serves as the normalization
anchor: measured $R(P)$ data (in arbitrary detected-count units) are scaled so
that their high-$\nin$ plateau (where the QD is fully saturated) approaches to
$\Rbare$. The low-power limit $R(\ninbeam\!\to\!0)$ is the coherent QD--cavity
response, and ($R(\ninbeam\!\to\!0)-\Rbare)/\Rbare$ is the reflectivity contrast. The Fock-space truncation is
chosen adaptively based on the number of photons inside the cavity.

\subsection{Pulsed second-order coherence}
\label{sec:defs}

The normally-ordered two-time second-order correlation of the detected field is
\begin{equation}
G^{(2)}(t,t+\tau) =
\expval{\adout(t)\,\adout(t+\tau)\,\aout(t+\tau)\,\aout(t)}.
\label{eq:G2_def}
\end{equation}
In a pulsed Hanbury Brown--Twiss measurement the coincidence histogram has a
peak at each multiple of the repetition period $\Trep$. The zero-delay
(``center'') peak measures photon pairs from the \emph{same} pulse; the peaks
at $\tau=m\Trep$ ($m\neq0$, ``side'') measure pairs from two \emph{independent}
pulses. The pulsed $\gtwo(0)$ is the center-peak area normalized by a side-peak
area,
\begin{equation}
{\;
\gtwo(0) = \frac{\mathcal{A}_{\mathrm{center}}}
{\mathcal{A}_{\mathrm{side}}}
= \frac{\int_{\mathrm{center}} g^{(2)}(\tau)\,d\tau}
{\int_{\mathrm{side}} g^{(2)}(\tau)\,d\tau}\; }
\label{eq:g2_ratio}
\end{equation}
where the emission-time-integrated, normalized correlation is
\begin{equation}
g^{(2)}(\tau) = \frac{G^{(2)}_{\mathrm{peak}}(\tau)}{\Nbar^{2}},
\qquad
G^{(2)}_{\mathrm{peak}}(\tau) = \int dt\; G^{(2)}(t,t+\tau),
\qquad
\Nbar = \int dt\, S(t).
\label{eq:peak_integral}
\end{equation}

This area-ratio construction is the
theoretical counterpart of integrating counts under each experimental peak and
normalizing by the uncorrelated side reference; it is insensitive to the peak
shape and to the overall detection efficiency. 
The quantum regression theorem is used to calculate $A_{center}$.

The side peak samples delays near $\tau=\Trep$. When $\Trep\gg\sigma_t,
\gpar^{-1}$, the two detection events lie in different pulses whose excitations
are statistically independent and the system relaxes fully between them, so
the correlator factorizes,
\begin{equation}
\expval{\adout(t)\,\adout(t+\Trep)\,\aout(t+\Trep)\,\aout(t)}
\longrightarrow S(t)\,S(t+\Trep),
\label{eq:factorize}
\end{equation}
requiring no regression. The side-peak shape versus the offset
$\delta\tau=\tau-\Trep$ is the autocorrelation of the single-pulse flux,
\begin{equation}
G^{(2)}_{\mathrm{side}}(\delta\tau) = \int dt\; S(t)\,S(t+\delta\tau)
= (S\star S)(\delta\tau).
\label{eq:autocorr}
\end{equation}
Normalizing by $\Nbar^2$ and integrating over the peak gives
$\mathcal{A}_{\mathrm{side}}$.

\subsection{Blinking model}
\label{sec:blinking}
Charge fluctuations intermittently drive the QD into a dark or charged
state, on timescales far longer than
the pulse-repetition period. On any given pulse the system is therefore either
in the neutral, optically active (``on'') configuration --- the coupled JC
system --- or in a blinked (``off'') configuration in which the QD is
off-resonant and the light sees the \emph{bare cavity}. Because blinking is
slow compared with the pulse train, the measured signal is a 
\emph{statistical mixture} of the two configurations, weighted by the fraction of
time $r$ spent in the neutral state~\cite{ReinhardVolz2012}.

If $R_{\mathrm{sim}}(\ninbeam)$ and $\gtwo_{\mathrm{sim}}(\ninbeam)$ be the coupled
(``on'') reflectivity and second-order coherence, and $\Rbare$ the bare (``off'')
reflectivity with $\gtwo_{\mathrm{off}}=1$ (coherent light), then the per-pulse
detected energies are
\begin{equation}
\Eon(\ninbeam) = R_{\mathrm{sim}}(\ninbeam)\,\ninbeam,
\qquad
\Eoff(\ninbeam) = \Rbare\,\ninbeam .
\label{eq:Eon_Eoff}
\end{equation}

Reflectivity is a first-order (mean-intensity) quantity, so the two
configurations add in proportion to their occupation to generate the observed reflectivity
\begin{equation}
{\;
R_{\mathrm{obs}}(\ninbeam) = r\,R_{\mathrm{sim}}(\ninbeam) + (1-r)\,\Rbare\; }
\label{eq:R_blink}
\end{equation}

Because $\gtwo$ counts photon pairs, each configuration contributes in
proportion to the \emph{square} of its detected energy. The mixed second-order
coherence is~\cite{ReinhardVolz2012}
\begin{equation}
{\;
\gtwo_{\mathrm{obs}}(\ninbeam) =
\frac{r\,\Eon^2\,\gtwo_{\mathrm{sim}}(\ninbeam)
+ (1-r)\,\Eoff^2\,\gtwo_{\mathrm{off}}}
{\big[\,r\,\Eon + (1-r)\,\Eoff\,\big]^2}\; }
\label{eq:g2_blink}
\end{equation}

